\documentclass[conference,a4paper]{IEEEtran}

\usepackage[acronym,nomain,nonumberlist]{glossaries}
\usepackage{glossaries}
\setacronymstyle{long-short}

\newacronym{sdr}{SDR}{software defined radio}
\newacronym{mmw}{mmWaves}{millimetre waves}
\newacronym{ota}{OTA}{over-the-air}
\newacronym{dut}{DUT}{device-under-test}
\newacronym{tf}{TF}{frequency transfer function}
\newacronym{ni}{NI}{National Instruments}
\newacronym{cir}{CIR}{channel impulse response} 
\newacronym{SNR}{SNR}{signal-to-noise ratio} 
\newacronym{los}{LOS}{line-of-sight}
\usepackage{siunitx}
\DeclareSIUnit{\sample}{S}

\usepackage{float}
\usepackage{xcolor}

\usepackage{cite}

\ifCLASSINFOpdf
  \usepackage[pdftex]{graphicx}
  \graphicspath{{../newfiguers/}{../}}
  \DeclareGraphicsExtensions{.pdf,.jpeg,.png}
\else
\fi
\begin{document}

\title{ A Bandwidth Scalable Millimetre Wave Over-The-Air Test System with Low Complexity}

\author{\IEEEauthorblockN{
Erich Z\"ochmann\IEEEauthorrefmark{1},
Terje Mathiesen\IEEEauthorrefmark{2}, 
Thomas Blazek\IEEEauthorrefmark{3},
Herbert Groll\IEEEauthorrefmark{3},
Golsa Ghiaasi\IEEEauthorrefmark{2}
}                                     
\IEEEauthorblockA{\IEEEauthorrefmark{1}
\textit{PIDSO - Propagation Ideas \& Solutions GmbH, Vienna, Austria,}}
\IEEEauthorblockA{\IEEEauthorrefmark{3}
\textit{Institute of Telecommunications, TU Wien, Vienna,Austria, }}
\IEEEauthorblockA{\IEEEauthorrefmark{2}
\textit{Department of Electronic Systems, NTNU, Trondheim, Norway, golsa.ghiassi@ntnu.no}}

}



\maketitle

\begin{abstract}
In this work, we show the design and validation of a testbed for \glsdesc{ota} testing of \glsdesc{mmw} equipments.  
We have extended the frequency capabilities of a baseband channel emulator which is capable of emulating non-stationary channels, to higher frequencies.  
To assure that the propagation between devices-under-test and the RF-frontend of the emulator is only a line-of-sight link, we have isolated the devices-under-test in two anechoic chambers. 
We characterize the testbed at \SI{57}{\giga\hertz} by means of frequency sweeps for two artificial cases:  when the emulator is replicating a one-tap channel and a two-tap channel. 
Finally we demonstrate the system functionality through reproduction of channels which were acquired during a vehicular millimetre wave  measurement campaign conducted in Vienna in 2018.

\end{abstract}


%

\vspace{7pt}
\section{Introduction}

5G communication systems rely on complex network topologies using new access technologies for communication links. With increased demand for (ultra) reliable communications envisioned in 5G, it is also vital to have repeatable tests which are capable of reconstructing complicated situational scenarios.  
The growing trend towards no-test on site, prompted to roll back the cost and logistic arrangement of testing, results in a growing interest in testbeds which evaluate the performance of physical devices in real-life scenarios with realistic communication links~\cite{golsa2018emulator}. 
These set-ups aim to fill the gap between network simulations with multiple nodes and on-field tests as they provide realistic real-time evaluation of physical radios in a given scenario.  
Real-time channel emulators are fundamental part of these testbeds as they replicate the physical characteristics of the propagation channel.  
At the simplest, realistic physical characteristics are achieved by playing back measured channels~\cite{blazek2018millimeter}.

 New frequency bands referred to as \gls{mmw} are soon deployed in order to achieve the ambitious 5G peak data rates ~\cite{roh2014millimeter}. Several \gls{mmw} measurement campaigns have been conducted recently~\cite{zochmann2018measured,groll2019sparsity,park2019multipath,park201828,wang2018fading,boban2019multi} and based on them
channel models applicable for channel emulation have been derived~\cite{zochmann2017two,TUW-270673,8612933,TUW-278530,blazek2018model,blazek2018approximating}.
At the new frequency bands, above $30$\,GHz, antenna arrays will be employed to overcome the high isotropic path loss~\cite{roh2014millimeter}. 
Connecting cables to all RF chains, i.e., conductive testing, is hence very challenging.  
Moreover, most 5G modems have antennas embedded on the board or on package which makes traditional conductive emulation prohibitive for these type of radios~\cite{zhang2009antenna}.  
Thus, a more general approach referred to as \gls{ota} emulation is required in order to enable integration of new modems into the testbeds~\cite{fan2018over}.

\subsection*{Contributions}

This work demonstrates a low complexity \gls{ota} set-up and investigates feasibility of emulating large bandwidths by employing several \glspl{sdr} in parallel. 
We show the \gls{tf} of the \gls{ota} chambers and the \glspl{tf} after inserting the \glspl{sdr}.
To verify our set-up, we emulate artificial one-tap and two-tap channels, and we further ``play-back'' a channel from a recent measurement campaign~\cite{groll2019sparsity}.

\begin{figure}
\centering
\includegraphics[width=0.49\textwidth]{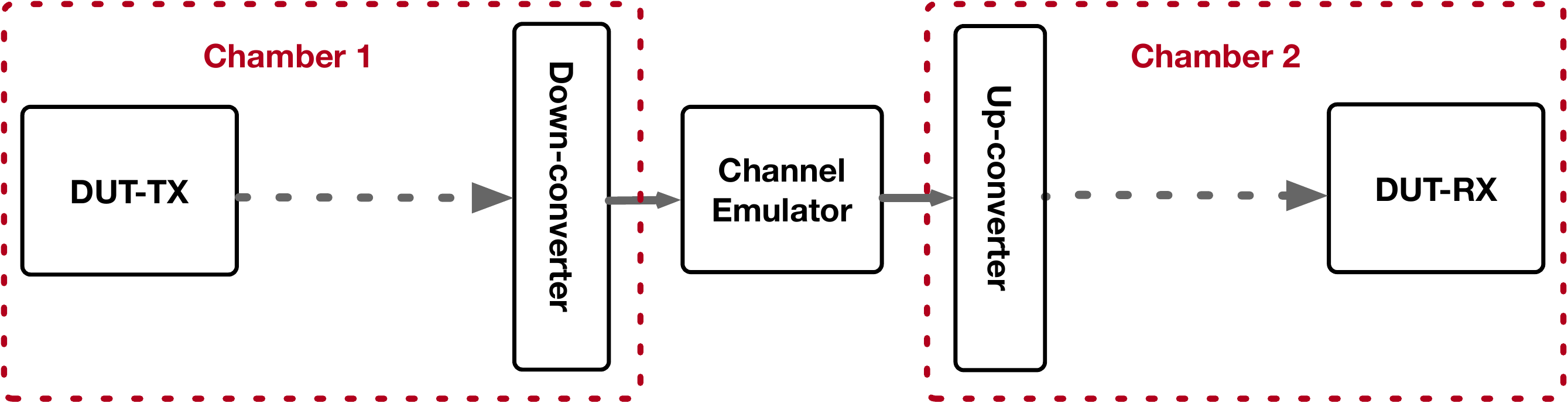}
\caption{Sketch of the OTA scheme. \label{fig:OTA_EM_schematic}}
\end{figure}

\vspace{7pt}
\section{OTA Chamber Measurements}

\begin{figure}
\centering
\includegraphics[width=0.49\textwidth]{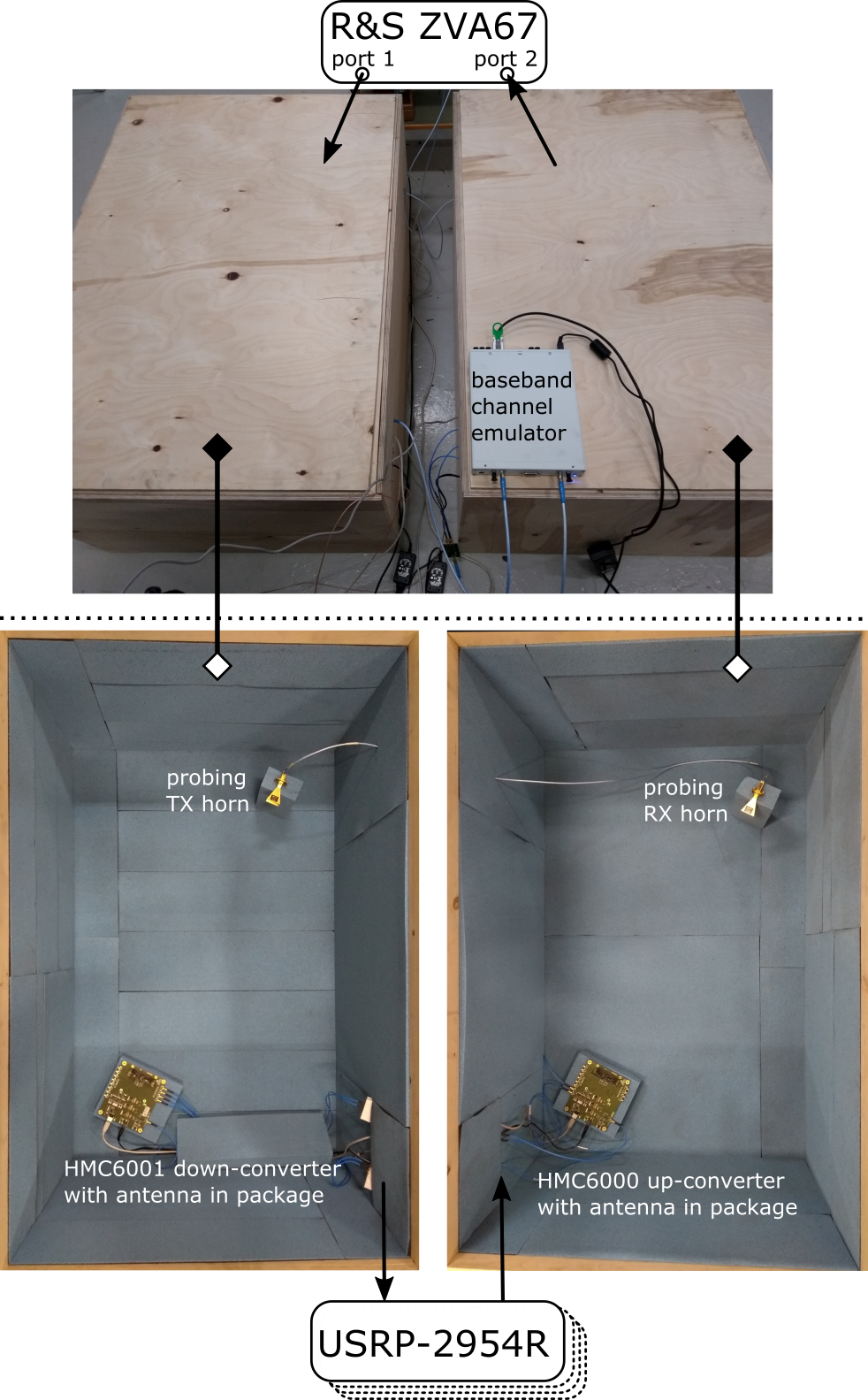}
\caption{OTA test set-up. For our measurements, a single USRP has been used sequentially. The full systems is envisioned to operate with several SDRs in parallel. \label{fig:OTAscheme}}
\end{figure}

\gls{ota} testing should occur in a controlled environment to ensure repeatability of the tests for comparison purposes, thus it is necessary to distinguish impacts of the test environment and of the channel to be replicated.  In the classic emulator testbeds, in order to achieve controlled environment, conductive material (such as cable) is used to establish the data links from the transmitter to the emulator and from the emulator to the receiver. When we have \gls{ota} links, 
we should ensure that the environment is well known and does not introduce frequency selectivity. Hence at the simplest it behaves like free space. 
To achieve this in our case, we isolate the emulator links from TX \gls{dut} antenna and to RX \gls{dut} antenna in two anechoic chambers. 

Foam based absorbent can be relatively thin for higher frequencies. We hence built our \gls{ota} chambers using approximately $4$\,cm thick absorbent.
The anechoic chambers are approximately $113\,\text{cm}\,\times\,73\,\text{cm}\,\times\,48\,\text{cm}\quad (L\,\times\,W\,\times\,H)$  in size.
These dimensions allow two antennas (with typical dimensions) to be separated by the far-field distance. For example, if we consider that a quadratic 8x8 $\lambda \slash 2$-distanced array at $60$\,GHz has a diagonal of less than $3$\,cm, then the Rayleigh distance will be at $36$\,cm. 

 Fig.~\ref{fig:OTA_EM_schematic} depicts the main idea behind the \gls{ota} system for testing equipments against replicated channels:  a \gls{dut} transmits its signal in the first anechoic chamber, where a down-converter shifts the \gls{ota} transmitted signal to baseband.
Then, a baseband emulator convolves the baseband signal with a (baseband) representation of the channel.
Finally, in a second chamber, the baseband signal is up-converted and transmitted to another \gls{dut} in receive mode. In this initial study, we replaced the \glspl{dut} with the vector-network analyser R\&S ZVA67. 
Our set-up is illustrated in Fig.~\ref{fig:OTAscheme}. 

Ideally, the whole down-conversion and up-conversion would be transparent as if a cable was used to connect the baseband ports and the set-up would be frequency flat. 

\begin{figure}
\centering
\includegraphics[width=0.42\textwidth]{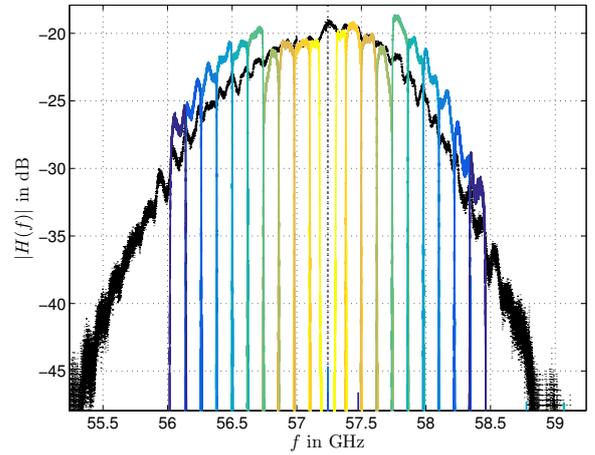}
\caption{Transfer functions of the chambers. The TF replacing the SDR (USRP-2954) by a cable is shown in black. The stitched versions by inserting the USRP and selecting a specific frequency band are shown in various colours.  \label{fig:TF}}
\end{figure}

\begin{figure}
\centering
\includegraphics[width=0.45\textwidth]{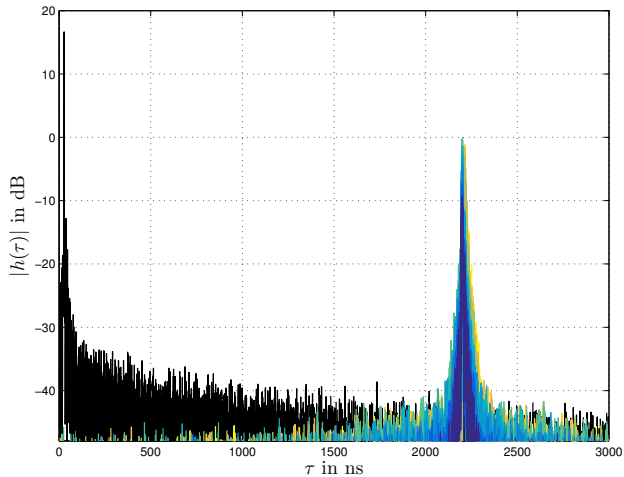}
\caption{Channel impulse response of the chambers. The \gls{cir} obtained by connecting the converters by a cable denoted in black has less than $100$\,ns initial delay. The frequency stitched channels show a distinct spike which is much broader due to band limitation. In addition, the baseband processing of the USRP introduces a delay of approximately $2.2\,\mu$s. \label{fig:CIR}}
\end{figure}

The measured \gls{tf} of two-chamber set-up connected through a cable is shown in black in Fig.~\ref{fig:TF}.
The frequency selectivity stems from the selectivity of the employed converters (HMC6000, HMC6001). 
Details of the chip-set used as RF frontend can be found in~\cite{zetterberg2015open}.
There is a ripple observed due to small reflections at connectors and baseband components and a pass-band behaviour resultant from the limited usable bandwidth of the converters.
The measured \gls{tf} is converted into the delay domain via the inverse discrete Fourier transform, see Fig.~\ref{fig:CIR}.
The obtained \gls{cir} shows the desired spike-like shape with almost negligible initial delay.
We hence conclude that the assembled anechoic chambers together with the utilized converters act as almost ideal one-tap channel, already without equalization.    

Next, we inserted an \gls{sdr}, namely, \gls{ni} USRP-2954R and emulated a single tap channel with \SI{120}{\mega\hertz} of bandwidth between the TX and RX chambers. We examined spectral stitching~\cite{shi2017246,dark2016spectral} by shifting the center of operation of the USRP by \SI{120}{\mega\hertz} at a time, and laying the acquired \glspl{tf} side by side.
It is observed that a small fraction of the possible bandwidth is cut out, due to the RF frontend of the USRP. 
 
\begin{figure}
\centering
\includegraphics[width=0.45\textwidth]{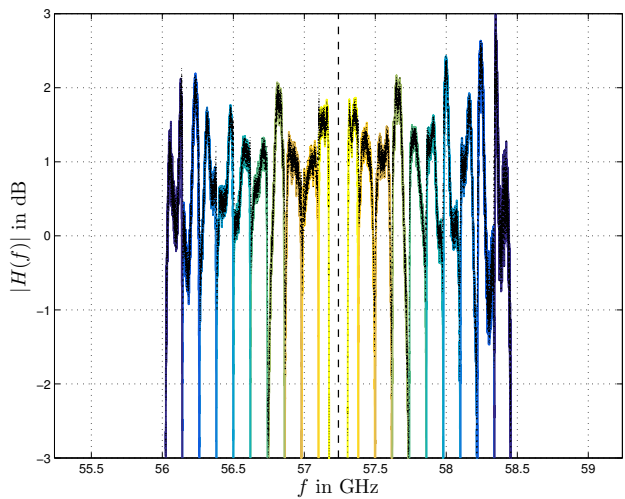}
\caption{Frequency transfer function after one tap equalization based on the mean amplitude. The remaining fluctuation are approximately $3$\,dB high. \label{fig:oneTap}}
\end{figure}

\begin{figure}
\centering
\includegraphics[width=0.45\textwidth]{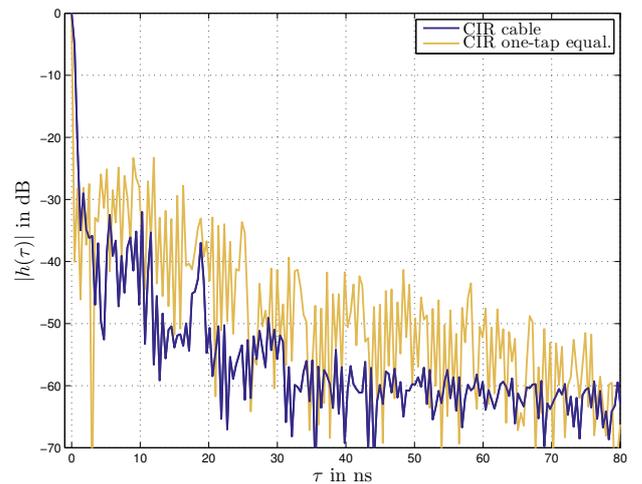}
\caption{CIR after one tap equalization compared to the cabled connection (without USRP). One tap equalization reduces the dynamic range of the emulator by 10\,dB as compared to the cable.  \label{fig:oneTapCIR}}
\end{figure}

The stitched sub-channels follow the frequency selectivity of the \gls{ota} set-up (chamber and up/down converters) and introduce no further selectivity, as shown in Fig.~\ref{fig:TF}.
At approximately $500$\,MHz the USRP changes the front-end, resulting in a higher gain.
This behaviour together with the overall frequency shape (low-pass shape) requires an equalizer when several USRPs operate on parallel sub-bands.
However, a simple one-tap equalizer (adapting the gain and the initial delay) suffices to obtain an almost frequency flat characteristic before channel emulation, see Fig.~\ref{fig:oneTap} and Fig.~\ref{fig:oneTapCIR}. 

%

\vspace{7pt}
\section{Emulated Channels}
The emulator has been implemented by running a custom design code on a \gls{ni} USRP 2954-R \gls{sdr} platform. The code is based on the work reported in \cite{golsa2018emulator} with modification to support the higher bandwidth of operation namely, \SI{120}{\mega\hertz}. The emulator implements a tapped delay line with 10 simultaneously active taps. The delay, and complex amplitude of each tap are updated for every channel snapshot in order to enable emulation of non-stationary channels. This allows us to aim for accurately replicating driving scenarios in which the channel dynamics change at a very fast rate.
 
\subsection{Two-Tap Channel}

To verify that our \gls{sdr} emulation is running properly, we emulate a simple two-tap channel~\cite{zochmann2017two} with equal gain for each tap, given as	
\begin{equation}
\label{eq: 2tap}
  h(\tau) = \delta(t)+ \delta(\tau-\tau_i)~.
\end{equation}

\begin{figure}[!h]
\centering
\includegraphics[width=0.4\textwidth]{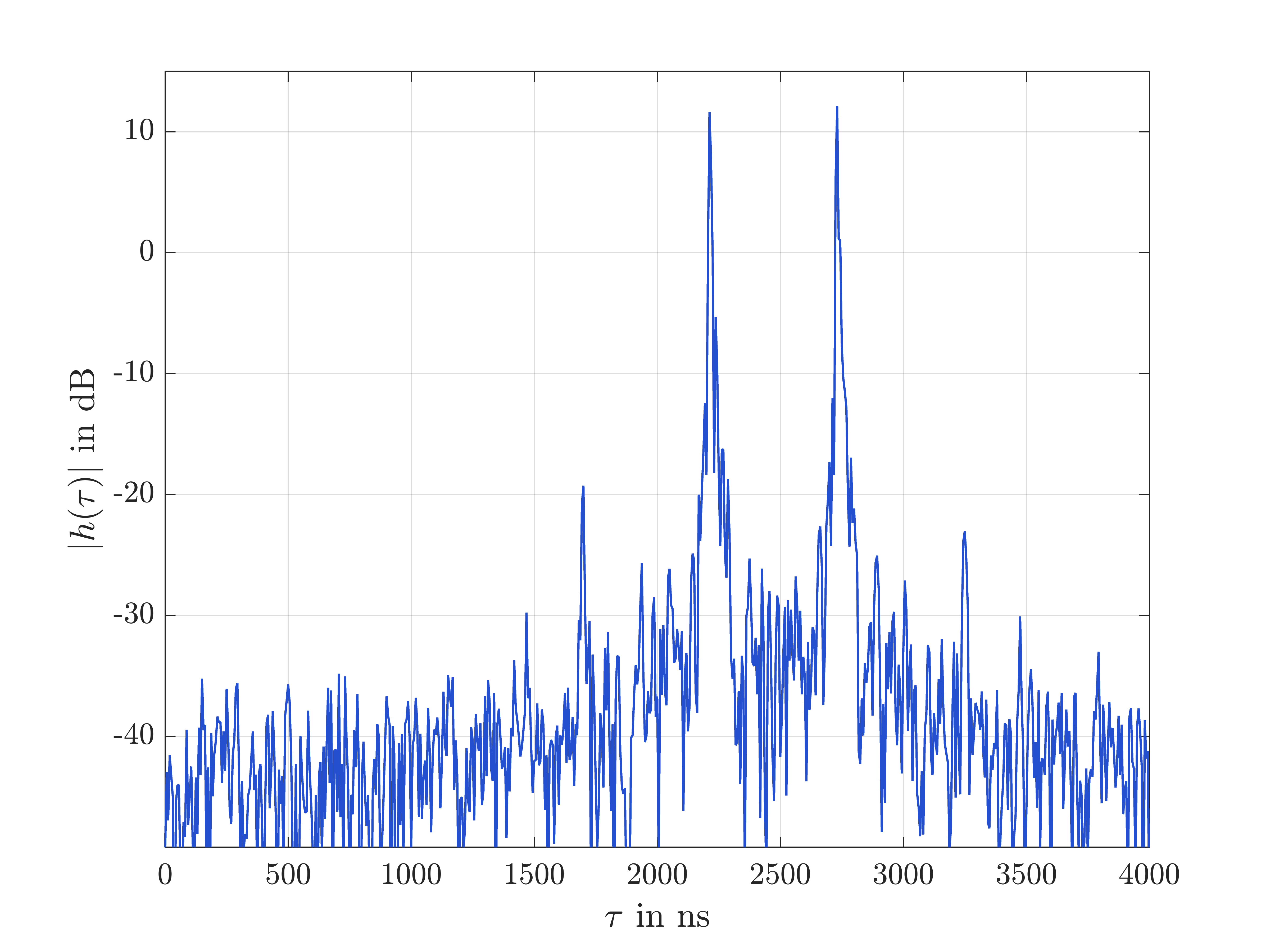}
\caption{CIR of the two-tap channel emulated in a single sub-band of $120$\,MHz. \label{fig:2tapCIR}}
\end{figure}

\begin{figure}[!h]
\centering
\includegraphics[width=0.4\textwidth]{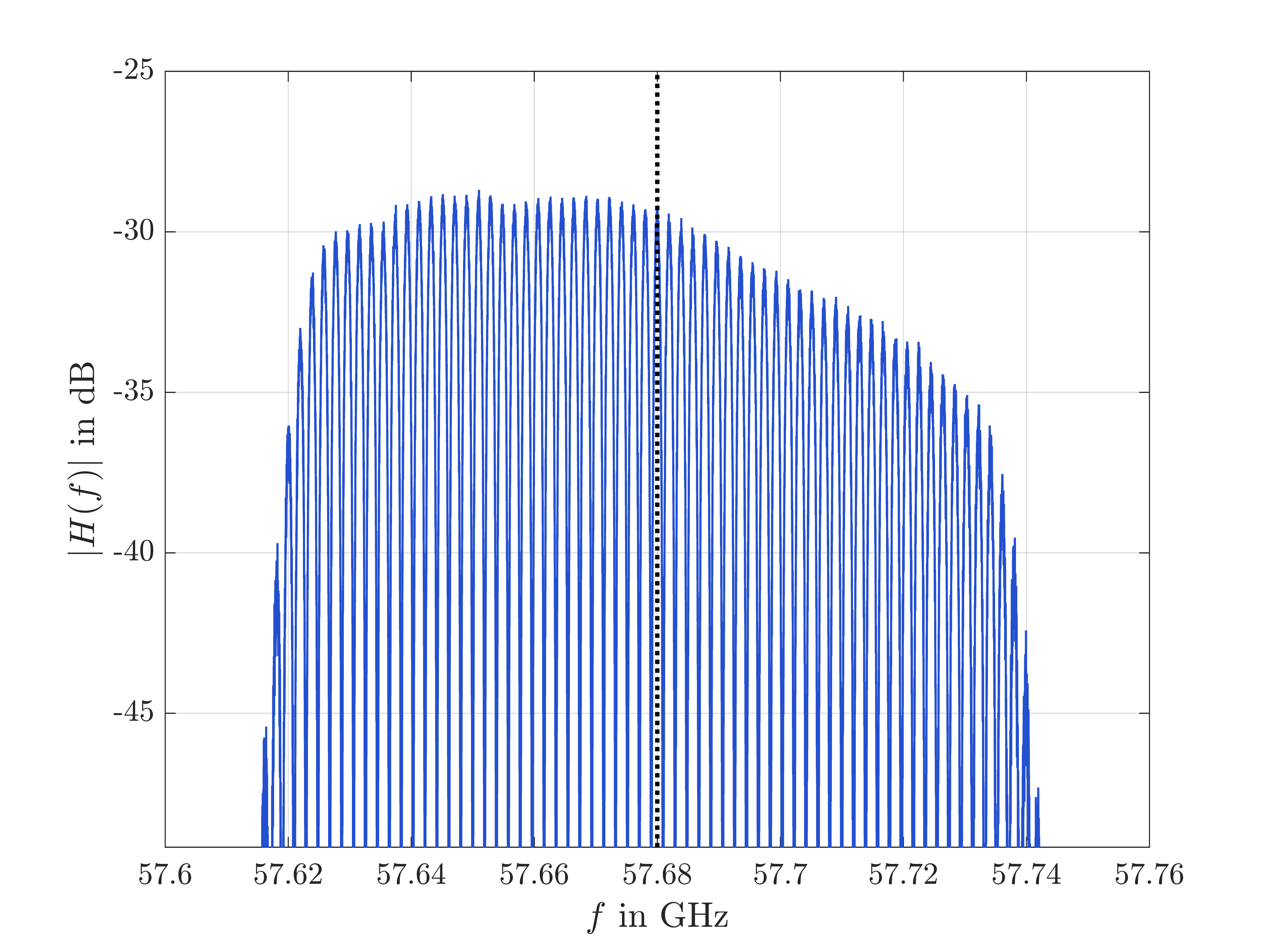}
\caption{TF of the two-tap channel emulated in a single sub-band of $120$\,MHz. \label{fig:2tapTF}}
\end{figure}

Figure~\ref{fig:2tapCIR} shows the first tap at the initial delay of approximately $2.2\,\mu$s and the second tap $500$\,ns later. This channel leads to a periodic pattern visible in the \gls{tf}, see~Fig.~\ref{fig:2tapTF}.
%

\subsection{Reproducing a Measured Channel}
\begin{figure}
    \centering
    \includegraphics[width=0.47\textwidth]{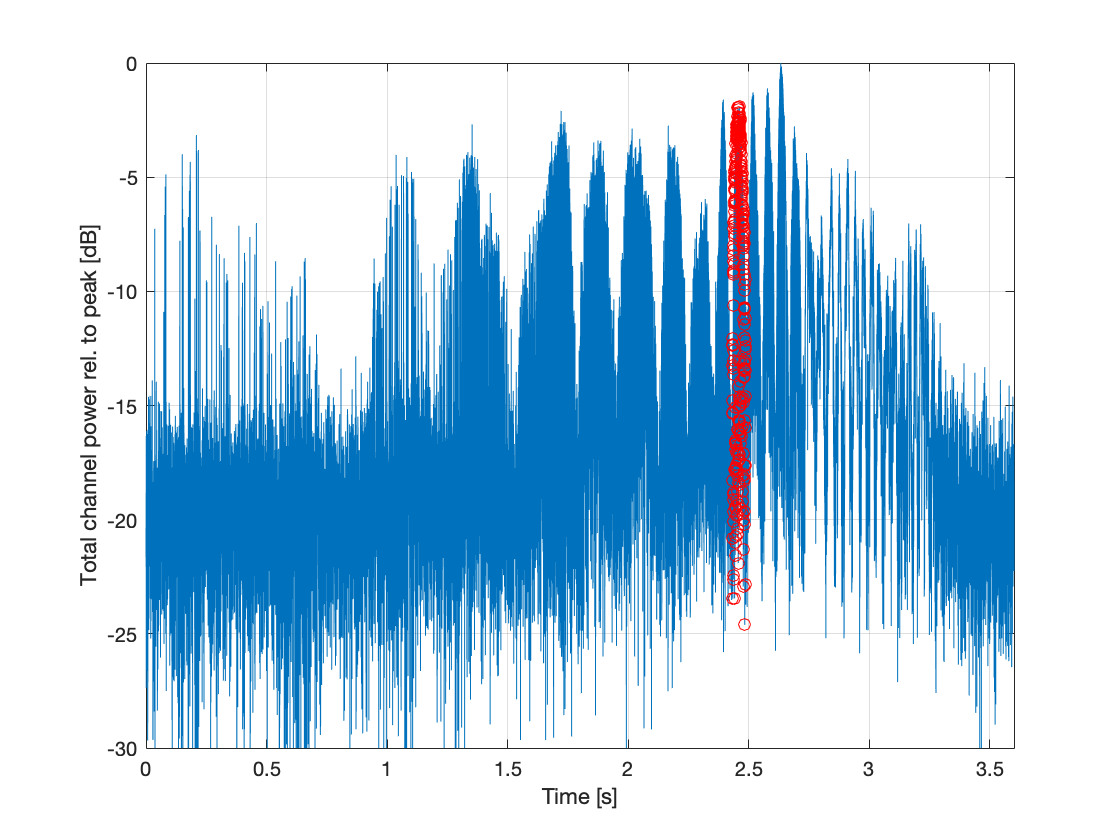}
    \caption{Total received power of the considered measurement. Full trace (blue) and considered region(red). Each circle denotes a snapshot of the measured channel.}
    \label{fig:measurement_trace}
\end{figure}
In this section, we demonstrate reproduction of a time-variant channel by ``play-back'' of the measurement campaign published in~\cite{groll2019sparsity}.
The measured scenario is V-band vehicle-to-infrastructure communications with a directive transmit antenna and an omni-directional receive antenna. 
A vehicle drives towards a street crossing at a speed close to $14$\,$\frac{\text{m}}{\text{s}}$ in an urban street canyon, while transmitting at $60$\,GHz with a bandwidth of $100$\,MHz with an horn antenna on the roof top directed in direction of driving.
The receiver is positioned directly at the crossing at $5$\,m elevation.
For this scenario, the received signal experiences distinct Doppler shifts between paths from the \gls{los}, the ground reflection, and multiple scatterers.
In~\cite{groll2019sparsity}, it has been shown that this channel is very sparse in the delay-Doppler domain, hence, the channel is approximated with $4$ active taps at a time.
The approximation follows the strategies proposed in~\cite{mecklenbrauker2017c,blazek2017sparse,zochmann2015density,blazek2018model}. 
As seen in \cite{groll2019sparsity}, the channel fluctuates semi-deterministically with the position. For the initial analysis to evaluate the emulator, we were interested in a section of the measurement that provided high \gls{SNR}. Hence, we focus on a section containing a total number of 301 snapshots starting at the $2.42$\,s mark for the duration of $53.5\,ms$, as shown in Fig.~\ref{fig:measurement_trace} by red marks. 

\begin{figure}
\centering
\includegraphics[width=0.47\textwidth]{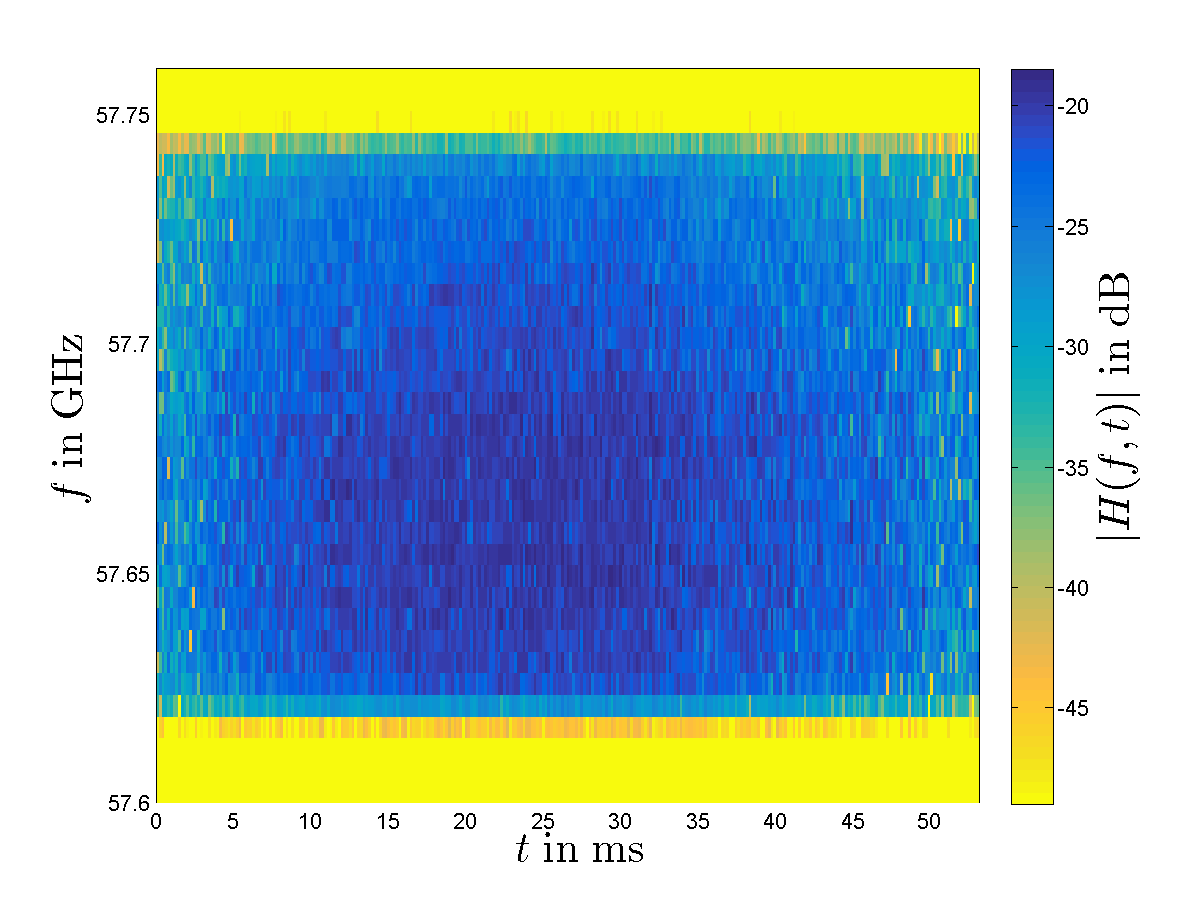}
\caption{Emulated time-variant transfer function of the measured channel. \label{fig:TVTF}}
\end{figure}

\begin{figure}
\centering
\includegraphics[width=0.47\textwidth]{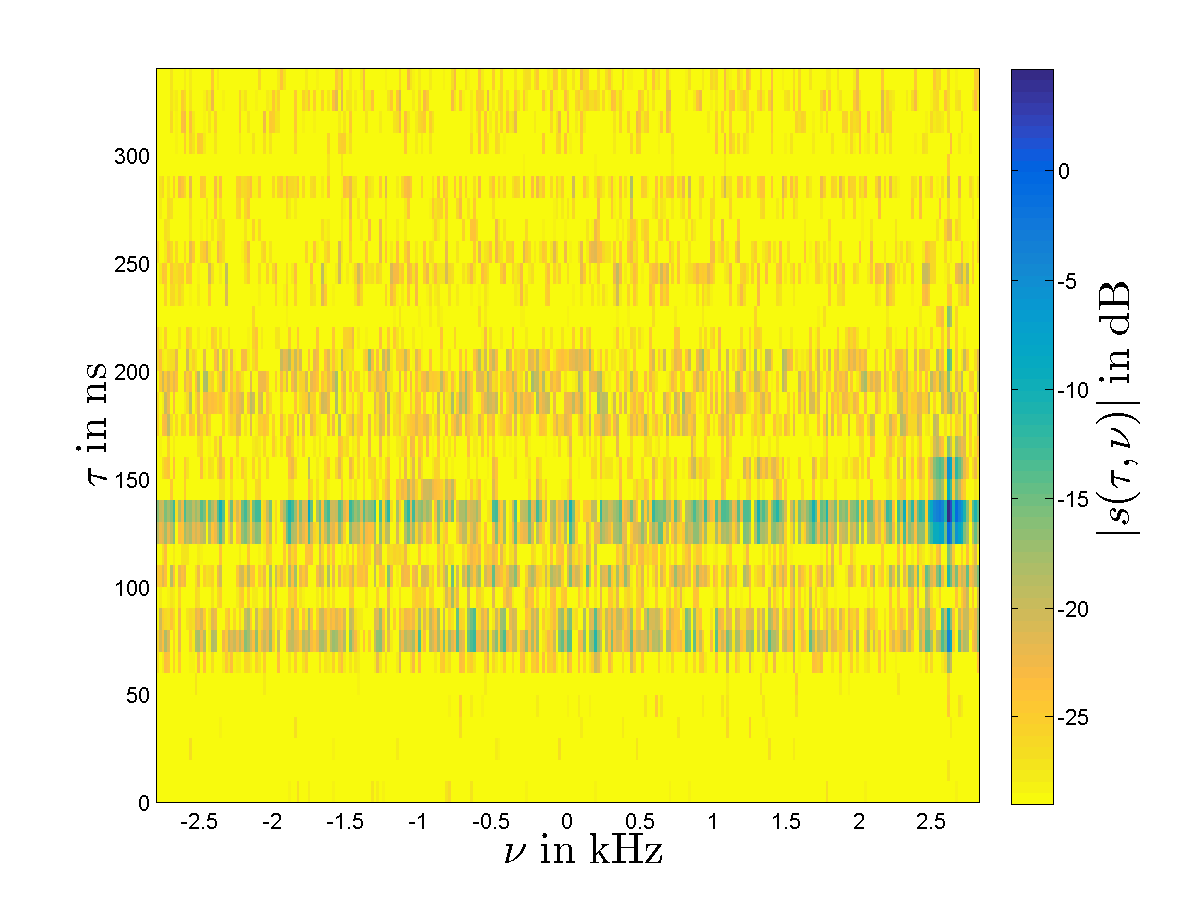}
\caption{Emulated delay-Doppler spreading function of the measured channel.
\label{fig:DDSF}}
\end{figure}

As the measurements were conducted with $100$\,MHz bandwidth, a direct emulation with a single USRP 2954-R is possible.
The time-variant \gls{tf}, its time-variant \gls{cir}, and its delay-Doppler spreading function are shown in Figs.~\ref{fig:TVTF}\,--\,\ref{fig:DDSF}, respectively. 
The respective formulas are found in~\cite{molisch2012wireless}.
\begin{figure}
\centering
\includegraphics[width=0.47\textwidth]{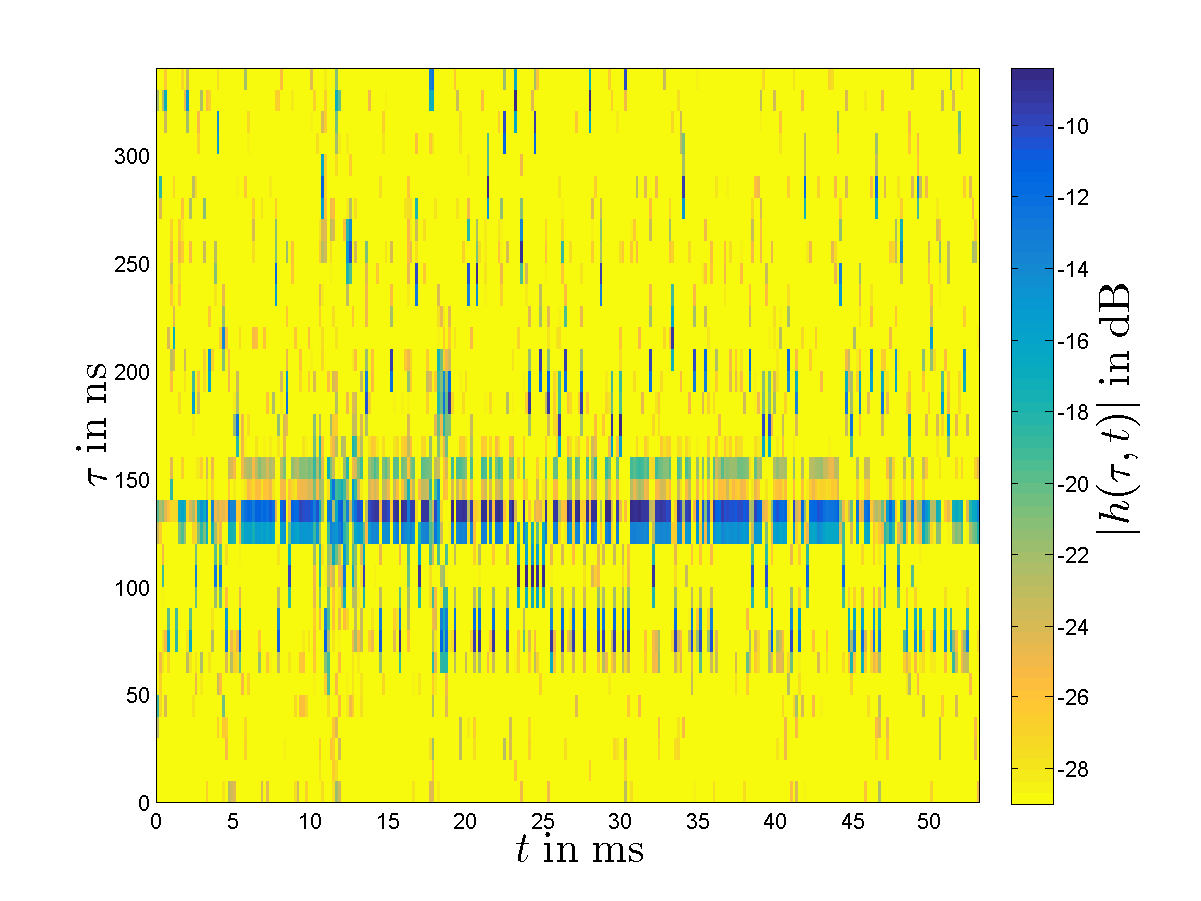}
\caption{Emulated time-variant channel impulse response of the measured channel.}
\end{figure}

Due to the relatively narrow bandwidth of $100$\,MHz, the time-variant frequency response does not show many notches in magnitude and appears relatively flat. The time-variant \gls{cir} is hence dominated by the \gls{los}.  Eventually, the delay-Doppler plane shows the dominant \gls{los} component at approximately $\nu=2.5$\,kHz as shown in Fig.~\ref{fig:DDSF}. The fades of the \gls{los} component are due to the street reflections which appear at the same delay tap.
The other fluctuations in either domain are due to noise originating from fitting of the channel to emulate 
and due to the non-ideal characteristics of the emulator.
\vspace{7pt}
\section{Conclusion}

This contribution shows a custom built millimetre wave over-the-air testing system with low complexity.
We show that our system is almost frequency flat and that multiband processing via several USRPs is feasible.
The wideband emulation only needs a one-tap equalization and appropriate channel representations at all subbands.
A mathematical formulation for selecting the channel taps at different sub-bands is already achieved~\cite{blazek2018approximating,blazek2018millimeter}.


\vspace{7pt}
\section*{Acknowledgment}

The authors would like to thank Christoph Friedrich Mecklenbr\"auker and Nils Torbj{\"o}rn Ekman for their insightful discussions. They are also grateful to Connectivity Technologies and Platforms department at Sintef Digital and ReRaNP lab at NTNU  for facilitating parts of the equipments. Special thanks goes to IES department at NTNU and IRACON COST action for their support to this project.



%
\vspace{7pt}

\bibliography{references}{}
\bibliographystyle{IEEEtran}

\end{document}